# An automated approach to extracting positive and negative clinical research results

Xuanyu Shi[1#], Shiyao Xie[2#], Wenjia Wang[3*], Ting Chen[4*], Jian Du[5*]

[1] *xyshi@stu.pku.edu.cn*,  [5] *dujian@bjmu.edu.cn*

National Institute of Health Data Science, Peking University, Beijing, 100191 (China)

[2] *xieshiyao@casisd.cn*,  [4] *chenting@casisd.cn*

Institutes of Science and Development, Chinese Academy of Sciences, Beijing, 100190 (China)

[3] *tsl-wangwenjia@tasly.com*

Cloudphar Pharmaceuticals Co., Ltd., Shenzhen, 518000 (China)

**Abstract**

Failure is common in clinical trials since the "successful" failures presented in negative results always indicate the ways that should not be taken. In this paper, we proposed an automated approach to extracting positive and negative clinical research results by introducing a PICOE (Population, Intervention, Comparation, Outcome, and Effect) framework to represent randomized controlled trials (RCT) reports, where E indicates the effect between a specific I and O. We developed a pipeline to extract and assign the corresponding statistical effect to a specific I-O pair from natural language RCT reports. The extraction models achieved a high degree of accuracy for ICO and E descriptive words extraction through two rounds of training. By defining a threshold of p-value, we find in all Covid-19 related intervention-outcomes pairs with statistical tests, negative results account for nearly 40%. We believe that this observation is noteworthy since they are extracted from the published literature, in which there is an inherent risk of reporting bias, preferring to report positive results rather than negative results. We provided a tool to systematically understand the current level of clinical evidence by distinguishing negative results from the positive results.

**Introduction**

Failure in science is an under-investigated area, although it is common especially in clinical trials (Yin, Wang, Evans, & Wang, 2019). According to the new evidence-based medicine pyramid (Murad, Asi, Alsawas, & Alahdab, 2016), randomized controlled trials (RCTs) are at the highest level while systematic reviews only play the role of lens inspecting all types of evidence. In other words, RCTs have the most reliable and straightforward conclusions when it comes to medical evidence appraisal. The results of RCTs are reported either in scientific publications or on registration platforms. Publication bias in RCTs is a noteworthy phenomenon that clinical researchers tend to publish statistically significant results over findings with no difference between the study groups (Easterbrook, Gopalan, Berlin, & Matthews, 1991), and it

---

has been causing time waste and false conclusions in the scientific community (Mlinarić, Horvat, & Šupak Smolčić, 2017). The discussions about failed and negative results have been gradually growing. Researchers proposed that negative results should be cared equally as positive results, and it is critical to publish all clinical trial studies regardless of the outcome (Nghiem, Vaidya, & Unger, 2022). In the area of cancer disease, negative trials contribute similar-sized scientific impact as positive trials, generating important observations on what new treatments should be avoided (Unger et al., 2016). Besides, questions to ask when a primary outcome fails were also discussed by researchers (Pocock & Stone, 2016), meaning failed outcomes can potentially have great values that should not be ignored. In other words, these are "successful" failures, since these negative results indicate the ways that should not be taken to treat cancer. It is a change of thinking from "to do what" to "not to do what". While encouraging researchers to disseminate synthesized analysis over positive and negative results, it is imperative for clinical researchers to automatically collect existing inconsistent results and structure them into a computational format. By distinguishing negative results from the positive results, both clinical researchers and public health policy-makers can capture a whole picture on both the successful and failed scientific evidence for informing an unbalanced and unbiased decision-making.

The Food and Drug Administration Amendments Act (FDAAA) in 2007 started to demand the studies registered on Clinicaltrials.gov to report results within 12 months of trial completion (Becker, Krumholz, Ben-Josef, & Ross, 2014). Meanwhile, abstracts in publications contain a huge number of results from RCTs. Compared to the structured and organized XML formatted results on the registration website, clinical trial results from peer-reviewed publications are represented in natural languages in abstract, which are strenuous for researchers to manually extract and curate. Trialstreamer (Marshall et al., 2020) is a database of over 800 thousand RCTs in a PICO (Population, Intervention, Comparation, and Outcome) format. It also has an extracted punchline sentence showing the results of each study. Limitations of Trialstreamer include 1) unmatched end-to-end I-O pairs, 2) non-standardized outcomes, and 3) lack of components representing positive and negative results, such as p-value, 95% CI (confidence interval) and other statistical effect size. In this study, we aim to solve the mentioned shortcomings by developing an integrated natural language processing pipeline with named entity recognition (NER) and relation extraction (RE) from RCTs results and conclusions. The proposed method focuses on the extraction of the statistical effects of interventions on outcomes. It is expected to help researchers acquire computational clinical evidence by discovering both positive and negative findings for medical knowledge mining, clinical decision support, systematic review, etc.

## Methods

The two core subtasks are the refined extraction of ICOE (E indicates the effect between an Intervention and an Outcome) and discovering the link between outcomes and effectiveness. Traditional approaches to named entity recognition and relation extraction use a pipeline approach to extract all entities mentioned in the entire abstract and then pass them to a relation extraction module (Lee Grace E, and, Sun Aixin 2019. Marshall IJ et al 2020. Nye BE et al

2021). This can lead to some unimportant information in the background presentation being extracted simultaneously as key information, which can be somewhat misleading in supporting subsequent in-depth analysis. In addition, we believe that information on the effectiveness of each outcome is essential information. Therefore, beyond the ICO entities, this paper focuses on the effectiveness extraction and the establishment of its association with the ICO entities.

These tasks may be viewed as instances of Sentence Classification combined with Named Entity Recognition (NER) and co-occurred-based relation extraction (RE), respectively. Motivated by observations concerning how results tend to be described in trial reports. We propose a new pattern-based ICOE extraction approach that firstly independently identifies sentences that describe interventions(I) and comparison(C) drug with grouping design, and extract I and C from the grouping design sentence. Secondly, identifying outcomes(O) and comparative effectiveness(E), then using the extracted information to discover their co-occurrence at the sentence level, and create a relationship between outcomes and comparative effectiveness.

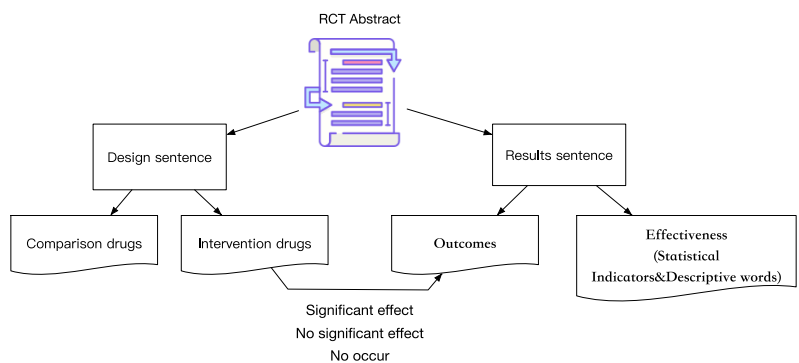

**Figure 1. Overview of ICOE recognition process**

The proposed extraction method is shown in Figure 2. The structural features of the abstracts are combined with the extraction of ICOE entities at the sentence level, thus optimizing the accuracy of the results. In RCT abstracts, there is an almost fixed pattern of writing some of the key aspects. The two most important sentences for extracting ICOE are the drug grouping design sentence and the outcome with comparative effectiveness sentence. The extract process in this paper is as follows:

    1) A sentence classification model was first trained to detect trail grouping design sentence. the design sentences describe drugs and population designs, for example *"This was a multicenter randomized controlled study including 96 patients with COVID- 19 who were randomly assigned into a **chloroquine (CQ)** group and a **favipiravir** group."* It can be seen that chloroquine (CQ) and favipiravir are the I and C entities, respectively. The linguistic and

textual features of this design sentence are obvious, usually including the words "patients / participants / randomized / assigned / group / versus etc", in addition to experimental protocol descriptions such as "*drug, dose (mg), frequency (two times a day), duration (for 14 days), etc*." A Bayesian classifier was used for sentence classification from the sentence level.

2) IC drugs extraction model was trained to obtain I and C entities from the detected drug grouping design sentence. Most studies have treated C as a subset of I. Although the relevant content is extracted by the model, the results may be inaccurate. In this paper we used a span-based NER model (Li et al., 2019) to learn from annotation data, I and C entities are extracted by identifying where they start and end in a sentence, respectively.

3) Afterwards, an Outcome NER model needed for obtain outcome named entity. The O named entity may be mentioned several times in the RCT abstract, and we are most interested in the O that is mentioned in the sentence where the effect is compared. For instance, "*one patient (2.3%) in the favipiravir group and two patients (4.2%) in the CQ group **died** (p = 1.00)*". The word died is the O entity in that sentence. A custom NLP model was trained to extract O entity from abstract level, the NLP framework spaCy3 was used for train the custom model based on annotated data.

4) After extracting the O entity, the next step is to extract the effectiveness(E) of the experiment results in the same sentence. There are usually two ways to describe the E in an RCT abstract: the first way is in the form of a statistical indicator, such as "**rate ratio, 0.95; 95% CI, 0.81 to 1.11; P-value= 0.50**". The other is in the form of a textual description, very often, authors will describe the effect of the outcome in natural language without any indicators, such as "**significantly higher, not different between, failed to attain statistical significance etc**". We therefore developed two ways of extracting the above.

  a) Statistical indicators: We divided the statistical indicators into two types, two templates were designed by using regular expressions for extracting indicators accurately. One type of indicators is the probability of outcome compared to controls group. Indicators include the hazard ratio (HR), odds ratio (OR) and relative risk (RR). It usually can be written in the following forms: *rate ratio, 0.95; 95% CI, 0.81 to 1.11; P-value= 0.50*. The template is: "INDICATOR NUM, NUM % CI NUM to NUM, P VALUE = NUM". The second type of indicator is a shorter version of the first, with only one indicator, the P-value, which is a direct comparison of the numerical differences in outcomes between the experimental and control groups. The template for second type indicator is: "P VALUE = NUM"

  b) Effect Description words: There are many effects description words that it is too time consuming to construct rules for extraction, thus we considered the effect description words extraction task as another special NER task, the same model used as O and training with different annotated data.

5) An Outcome NER model and a results descriptive words NER model were trained to obtain both outcome entity and descriptive words entity, a rule-based results indicator extraction method was designed for extracting effectiveness indicator. We have set a rule if outcome entity and results indicator or descriptive words co-occurrence in some sentence, the sentence is a clinical outcome results description sentence.

6) Last step is correlating the O and E by using sentence level co-occurrence relationship. If O and E co-occurred in same sentence, most likely E is O's results of effectiveness.

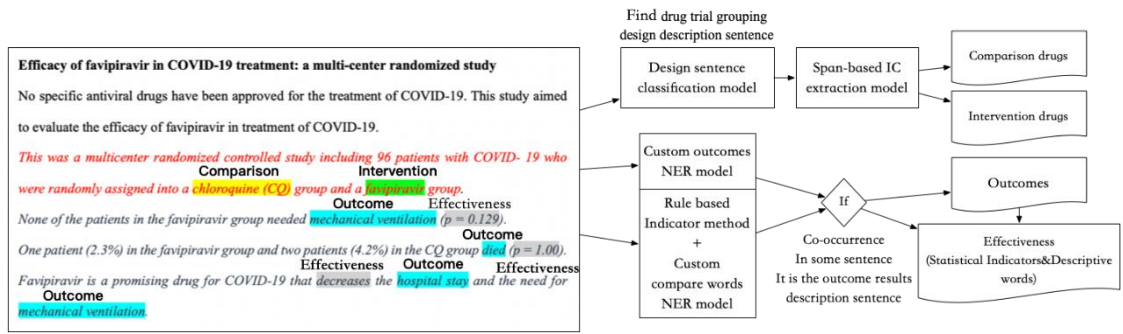

Figure 2. Overview of abstract pattern based ICOE extraction method

**Data**

We downloaded 216 abstracts of COVID-19 RCT studies from MEDLINE database. One medical professional designed the annotation rules and then two researchers from same medical college finished the annotation process. We cross-validate the annotation results, and the disputed results are finally judged. Four classes of entities in abstracts have been labeled: I, C, O and E- Description words. We don't annotate the E-indicators such as "odds ratio", "95% CI" and "p-value" because they are highly formalized and can be extracted by regular expression easily. Then we split this dataset into train set (80%) and test set (20%). A RCT abstract usually contains one intervention/comparison but a few types of outcomes and corresponding compare words. We count the number of unique entities in each data set. The 216 annotate data is our gold standards.

216 annotated data for training a valid model is not enough, so we collected all COVID-19 RCT studies in PubMed database, after the removing non-drug treatment studies, there were 836 remaining studies. Instead of manual annotation, labels in the larger dataset were semi-automatically generated using the predictions of the model trained on the gold standards dataset, after a quick manual screening of the incorrect predictions. The statistical result of dataset is shown on Table 1.

**Table 1. Count of annotated data**

|  | Gold standards | Semi-automatically generated data |
|---|---|---|
| RCT Abstracts | 216 | 836 |
| Labeled Sentence for classification | 216 | 492 |
| Labeled I/C entities | 178 | 477 |
| Labeled Outcome entities | 613 | 1476 |
| Labeled Compare word entities | 47 | 171 |

**Results**

(1) Performance of the extraction models

For evaluation purpose, 5-fold cross-validation is applied. data are equally divided into 6 groups. Among each run of model learning and testing, 4/5 of the data used as training set, 1/5 as test set. We report the performance on test sets. The model performance is reported based on the test set in Table 2. We test performance with different training data (Gold standards / Semi-automatically generated data) and different Entities. For each model setting, we report the best evaluation among 6 sets for cross-validation and also averaged measures, the classic evaluation metrics F1 score was used measure performance.

Table 2. Models' F1 score in different training sets

|  | Gold standards | Semi-automatically generated data |
|---|---|---|
| I/C | 0.866 | 0.905 |
| O | 0.835 | 0.932 |
| E-Description words | 0.701 | 0.814 |

As can be seen from the table, the models achieve a high degree of accuracy for ICO and E-Description words extraction through two rounds of training. A sample recognition result is shown in table3, An RCT abstract can be transformed into a set of ICOE entities, while including the correspondence between O and E. As can be seen in the sample, I is Favipiravir, C is Standard care alone. There are five outcomes, four of which correspond to E as a statistical indicator and one of which corresponds to E as a textual description.

It is worth noting that usually only one drug is tested in a paper, so the relationship between I and O is naturally formed in output result. Thus, ICOEs can eventually be linked as a small knowledge graph. More applications can be attempted in the future based on the knowledge graphs.

Table 3. Sample output for our ICOE extraction method

| PMID：34849615 | I/C | O-E |
|---|---|---|
| Efficacy of Early Treatment With Favipiravir on Disease Progression Among High-Risk Patients With Coronavirus Disease 2019 (COVID-19): A Randomized, Open-Label Clinical Trial [abstract image] | **I:** Favipiravir **C:** Standard care alone | **O:** Clinical progression to hypoxia  **E:** odds ratio, 1.30; 95% CI: .81-2.09; P-value=.28 |
| | | **O:** Mechanical ventilation  **E:** odds ratio,1.20;95% CI:.36-4.23;P-value=.76 |
| | | **O:** ICU admission  **E:** odds ratio, 1.09; 95% CI: .48-2.47; P-value=.84 |
| | | **O:** risk of disease progression  **E:** odds ratio, 12.54; 95% CI: .76-207.84; P-value=.08 |
| | | **O:** disease progression  **E:** not prevent |

(2) Positive versus Negative Results

We extracted P-values separately and analyzed the percentage of positive versus negative results. We use the traditional statistical P-value threshold of 0.05 to distinguish between positive and negative results. Results with P-value <0.05 represent the intervention group (I) has significant difference on the outcome (O) compared to the comparison group (C). Oppositely, results with P-value >= 0.05 represent the intervention group (I) has no significant difference on the outcome (O) compared to the comparison group (C).

The language expressions for P-values are difference across study results: The following table shows the counts of each expression and corresponding positive/negative mapping. The positive results accounted for 62% of all results while 38% are negative. Table 4 shows the full statistics of positive and negative P-values with different operators.

**Table 4. The statistics of positive and negative results by P-value**

| Operator | Count | Positive | Negative |
| --- | --- | --- | --- |
| Equal to (=) | 777 | 417 | 360 |
| Greater than (>) | 31 | 0 | 31 |
| Less than (<) | 219 | 217 | 2 |
| Greater than or equal to (>=) | 0 | 0 | 0 |
| Less than or equal to (<=) | 8 | 8 | 0 |
| **Total** | **1035** | **642 (62.0%)** | **393 (38.0%)** |

## Conclusions

To the best of our knowledge, this is the first efforts to extract the statistical effect of a specific intervention-outcome pair from natural language RCT reports in which multiple outcomes measured are described in one sentence. By defining a threshold of p-value, we find in all Covid-19 related intervention-outcomes pairs with statistical tests, negative results account for nearly 40%. We believe that this observation is noteworthy since they are extracted from the published literature, in which there is an inherent risk of reporting bias, preferring to report positive results rather than negative results. Medical bibliographic databases and clinical trial registries are two main sources of RCT reports. Compared with the former one, a registry like ClinicalTrials.gov consists of not only more than four hundred thousand structured and semi-structured trial information but also more than fifty-four thousand reported results. We could estimate that if the results reported on Clinicaltrials.gov are included, the proportion of negative results will further increase. Next, we will continue to detect the conflicting intervention-outcome pairs and analyze the reasons, such as different conditions or sample size. In addition, we also need to continue to improve the definition of positive results and negative results, beyond p-value by introducing "95% confidence interval" and other statistical effect indicators. Through a critical scrutinization on what is known, clinical researchers can systematically understand the current level of scientific evidence. It is also in accordance with the idea of metaknowledge (i.e., knowledge on knowledge), which will enable researchers to reshape science—to identify areas in need of reexamination and reweight former certainties (Evans & Foster, 2011)

## Acknowledgments

This work is funded by the National Key R&D Program for Young Scientists (2022YFF0712000), the National Natural Science Foundation of China (72074006), and